\documentclass[%
 aip,
 jmp,%
 amsmath,amssymb,
 reprint,%
author-numerical,%
nofootinbib,superscriptaddress]{revtex4-1}

\usepackage{graphicx}
\usepackage{dcolumn}
\usepackage{bm}
\usepackage{subfigure}
\usepackage{color} 

\begin{document}

\preprint{AIP/123-QED}

\title[]{A New Scheme for High-Intensity Laser-Driven Electron Acceleration in a Plasma}

\author{S. P. Sadykova}%
\altaffiliation{Electronic mail: \textcolor{blue}{Corresponding author - s.sadykova@fz-juelich.de}}
\affiliation{Forschungszentrum Julich, J{\"u}lich Supercomputing Center, J{\"u}lich, Germany
}%
\author{A. A. Rukhadze}
\email{rukh@fpl.gpi.ru}
\affiliation{Prokhorov General Physics Institute,
Russian Academy of Sciences, Vavilov Str. 38., Moscow, 119991, Russia}
\author{T. G. Samkharadze}
\affiliation{Moscow State University of Instrument Engineering 
and Computer Science
Moscow, 107996, Stromynka str. 20,  Russia} 




\date{\today}
\begin{abstract}
We propose a new approach to high-intensity relativistic laser-driven electron acceleration in a plasma. Here, we demonstrate that a plasma wave generated by a stimulated forward-scattering of an incident laser pulse can be in the longest acceleration phase with injected relativistic beam electrons. This is  why the plasma wave has the maximum amplification coefficient which is determined by the acceleration time and the breakdown (overturn) electric field in which the acceleration of the injected beam electrons occurs. We must note that for the longest acceleration phase the relativity of the injected beam electrons plays a crucial role in our scheme. We estimate qualitatively the acceleration parameters of relativistic electrons in  the field of a plasma wave generated at the stimulated forward-scattering of a high-intensity laser pulse in a plasma.

\end{abstract}

														
\maketitle

\begin{figure}
\centering
\includegraphics[width=0.7\linewidth]{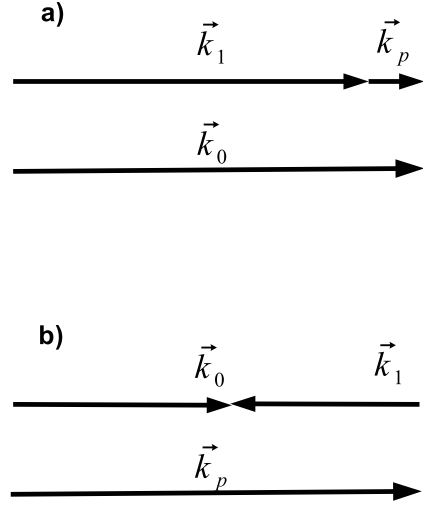}
\caption{Schematic illustration of a wave vector model for interaction of the high-intensity laser pulse ($\vec k_0$) with plasma which generates a plasma wave ($\vec k_p$) and forward- ($\vec k_1$) (a) and backward-scattered ($\vec k_1$) (b) waves, where  $k_0\simeq k_1>>k_p$  is valid for the  forward-scattering case, and $\vert k_0\vert \simeq \vert -k_1\vert<<\vert k_p \vert$ - for the back- scattering case .}
\label{Fig:1}
\end{figure}

 During the past few decades plasma accelerators have attracted increasing interest of scientists from all over the world due to its compactness, much cheaper construction costs compared to those for conventional one and various applications ranging from  high energy physics to medical and industrial applications. An intense electromagnetic pulse can create a plasma oscillations through  the stimulated scattering. Relativistic electrons injected into the plasma wave can be accelerated to much higher energy than nonrelativistic one.\\
\indent  The idea to accelerate the charged particles in a plasma medium using collective plasma wave fields generated by the high-energy electron beams belongs to the Soviet physicists G. I. Budker, V. I.  Veksler and Ia. B. Fainberg \cite{Budker,Veksler,Fainberg} in 1956, whereas assumptions for  generation of plasma  Langmuir waves by nonrelativistic electron bunches propagating through plasma were first made earlier in 1949 \cite{1, 2}.  High-energy bunch electrons generate a plasma wave in such a way that the energy from a bunch of electrons is transferred to the plasma wave through stimulated Cherenkov resonance radiation producing high acceleration electric fields. Later on, another acceleration scheme using a laser \cite{Dawson2} or time-shifted sequence of  bunched high-energy electrons injected into a cold plasma was proposed \cite{Dawson}.  In recent experiments at the Stanford Linear Accelerator Center it was shown that an energy gain of more  than 42 GeV was achieved in a meter long plasma wakefield accelerator, driven by a 42 GeV electron beam \cite{Blum}. For a detailed review about the modern status of this  research  field we would like to refer a reader to \cite{Joshi, Liver}.  \\
\indent Early experiments of the 60s and 70s demonstrated that efficiency of acceleration using the high-energy beams is much less than the expected one and the  generated field is much lower than a breakdown (overturn) electric field \cite{Akh}:

\begin{equation}\label{4}
{E_p}_{max}=\sqrt{2} m V_p\omega_p/e.
\end{equation}
where $e$- electron charge, $m$ - its mass, $V_p$ - plasma wave phase velocity, $V_p=\omega_p/k_p $ where $k_p$ - plasma wave vector, $\omega_p$ - plasma frequency, ${\omega_{p}}=\sqrt{4\pi e^2 n_{e}/m}$ with $n_e$ being the electron density and m - its mass. The explanation for the experiments failure was given in the work \cite{9} where it was shown that trapping of electrons in a generated by a beam plasma wave occurs and as a result growth of the field amplitude stops when the field amplitude is much less than that of the breakdown field \cite{4, 5}.\\
\indent With the appearance of the high-intensity lasers in the 80s, a new era of plasma acceleration has begun. The hopes were fed by  possibility of the stimulated scattering of a laser pulse by plasma electrons in a rare plasma with the generation of high-intensity  longitudinal plasma wave where the electron trapping does not occur and, as a result, high fields with respect to the breakdown field  ${E_p}_{max}$ can be generated. However, there are still a lot of unsolved problems related to the development of instabilities hindering the laser-driven plasma-based acceleration \cite{Liver}.   \\
\indent The problem of plasma stability in external nonrelativistic high-frequency field (laser) $E_0(t)=E_0 \sin(\omega_0 t+\vec{k}_0 \vec{r})$ with $E_0$ being the amplitude and $\omega_0$ being the frequency of the incident laser  wave ($\omega_0 >> \omega_p$) was extensively studied in \cite{4,5}. The scattering of a high-frequency transverse electromagnetic wave in a plasma  which excites the longitudinal plasma wave can be described by the following dispersion relation \cite{4}:
\begin{equation}
\label{30}
\left({\omega}^2-{\omega_p}^2\right)\left({\omega_s}^2-c^2k_s^2\right)=\frac{{\omega_p}^2k^2{V_E}^2}{4},
\end{equation}
where $\omega_s=\omega_0-\omega$ - transverse scattered wave frequency, $\vec{k}_s=\vec{k}_0-\vec{k}$ - wave vector of the scattered wave, $\omega=\omega_p+i\delta$ with $\delta$ being the increment, $\omega$ and $\vec k$ are the frequency and wave vector of a generated plasma wave respectively, $\omega_p >>k v_{Te}, \omega_i$,  where $v_{Te}$ - electron thermal velocity, $\omega_i$ - ion frequency, and $V_E=eE_0/m\omega_0$ ($V_E<< c$) - velocity of plasma electrons oscillating in a laser field. Herefrom,  two limiting cases follow: 1) The Raman forward-scattering (towards the injection of a laser pulse) with the increment $\delta_f=(1/4)\omega_p({V_E}^2\omega_p /c^2\omega_0)^{1/2}$, 2) The Thomson back-scattering (backwards the laser pulse) with the increment $\delta_b=(\sqrt{3}/2)\omega_p({V_E}^2\omega_0/(2c^2\omega_p))^{1/3}$ which is much higher than $\delta_f$. In this case the wake is generated which is well studied in many works starting from \cite{Fainberg} till the modern days \cite{Liver}.\\
	\indent A fundamental issue of conventional acceleration schemes employing parametrically driven plasma wave generation is the following: in spite of the fact that the instability induced by the backward-scattering  generating a plasma wave or a wake has a maximum increment $\delta_b$ compared to that generated by a forward  one $\delta_f$ by $3(\omega_0/\omega_p)^{5/6} $ times \cite{4} (see Fig. \ref{Fig:1}), this acceleration scheme is not suitable for particle acceleration because such lasers have a very short laser pulse length. Since the wave vector of a plasma wave $\vec k_p$ is equal to the double magnitude of that of a laser pulse ($k_p \simeq 2\omega_0/c$), the phase velocity of a plasma wave is quite low. Due to this fact the wave leaves behind both the laser and the backward-scattered waves getting localized at the back to the front of the laser pulse. 
	As a result the plasma wave gets soon out of the acceleration phase with the laser wave and the electron beam injected into the plasma gets soon out of a phase with the plasma wave what halts the acceleration process. \\
	\indent	In this work, we would like to propose another acceleration scheme, namely: a stimulated forward-scattering based plasma acceleration of relativistic electrons. Due to the stimulated laser forward-scattering a  plasma wave is generated as well. In this case, the plasma wave and the injected electrons can stay in acceleration phase for a much longer time. The dispersion relation (\ref{30}) can be easily generalized for relativistic laser pulse when $V_E \simeq c$. This can be done only for a laser with circular polarization when the high frequency harmonics are absent \cite{Mark}. In this case, Eq. (\ref{30}) takes the following form:
	\begin{equation}
\label{301}
\left({\omega}^2-\frac{{\omega_p}^2}{{\gamma_0}}\right)\left({\omega_s}^2-c^2k_s^2\right)=\frac{{\omega_p}^2k^2{V_E}^2}{4{\gamma_0}^3},
\end{equation}
where $\gamma_0=(1-{V_E}^2/c^2)^{-1/2}$ ($V_E \to c$). From this equation follows that the instability increment increases proportionally to $V_E$ growth when $V_E<<c$, and - decreases when $V_E\to c$. The corresponding increment  magnitudes are for the forward Raman scattering: $\delta_f \sim\omega_p({V_E}^2\omega_p /c^2\omega_0\cdot (1/{\gamma_0}^{7/2}))^{1/2}$ and for the Thomson scattering: $\delta_b \sim\omega_p({V_E}^2\omega_0/(2c^2\omega_p)\cdot (1/{\gamma_0}^{3}))^{1/3}$ and $\delta_f/\delta_b <<1$ by $3(\omega_p/\omega_0)^{5/6} (V_E/{\gamma_0}^{9/4} c)^{1/3})$ times. Their maximum increment magnitudes respectively: $\delta_f \sim 0.4\omega_p(\omega_p /\omega_0)^{1/2}$ at $V_E/c\simeq 0.6$ and $\delta_b \sim 0.56\omega_p(\omega_0/(2\omega_p))^{1/3}$ at $V_E/c\simeq 0.63$. \\
Correspondingly, the breakdown electric field will turn into \cite{Akh}:
\begin{equation}\label{44}
{E_p}_{max}=\sqrt{2} m V_p\omega_p\sqrt{\gamma_0}/e,
\end{equation}
where $V_p\simeq c$.\\
\indent According to our studies, we must note that the longest acceleration time takes place only when the injected electrons are relativistic compared to the earlier known works in such research field described in \cite{Liver} where the electrons were either nonrelativistic or weakly relativistic. We presume that this factor assisted in the successful acceleration made in \cite{Ross}.   \\    
Let the laser pulse with Langmuir frequency ${\omega_{0}}$ be injected into the cold plasma at  $\boldsymbol{{\omega_0}^2>>{\omega_p}^2}$. Consider the case when  $Z||\vec{V_0}$, $\vec{V_0}$, $\vec{V_1}$ and $\vec{V_p}$ - phase velocities of laser, forward-scattered and plasma waves.
Here, we employ the CGS system of units. The parametric resonance for the stimulated forward-scattering can be written as following:
\begin{equation}\label{1}
\begin{gathered}
\omega_0=\omega_1+\frac{\omega_p}{\sqrt\gamma_0} \hfill \\
k_0=k_1+k_p,
\end{gathered}
\end{equation}
where $k_0$, $k_1$ and $k_p$ are the wave vectors of the incident laser pulse, forward-scattered wave and plasma wave respectively, $\omega_1$ and $\omega_p$ - corresponding real parts of the  frequencies $\omega_s$ and $\omega$: 
\begin{equation}\label{2}
\begin{gathered}
\omega_0=\sqrt{\frac{\omega_p^2}{\gamma_0}+k_0^2 c^2} \hfill \\
\omega_1=\sqrt{\frac{\omega_p^2}{\gamma_0}+k_1^2 c^2},
\end{gathered}
\end{equation}
After having solved the system of Eqs. (\ref{1}) and (\ref{2}) one can find out that the phase velocities of three waves are the following
\begin{equation}
\label{3}
\begin{gathered}
\left(V_0=\frac{\omega_0}{k_0}\right)=\left(V_1=\frac{\omega_1}{k_1}\right)=c\left(1+\frac{\omega_p^2}{2\gamma_0\omega_0^2}\right), \hfill \\
\left(V_p=\frac{\omega_p}{\sqrt{\gamma_0} k_p}\right)=c\left(1-\frac{\omega_p^2}{2\gamma_0\omega_0^2}\right)
\end{gathered}
\end{equation}
where $V_0=V_1>c$, $V_p<c$ and approximately equal to each other $V_0\simeq V_1 \simeq V_p$ since ${\omega_p}^2/{\gamma_0\omega_0}^2<<1$.
These phase velocities of incident and scattered waves are slightly higher than the speed of light,  whereas that of a plasma wave is slightly less like in a laser-driven wakefield regime described in \cite{Gibbon}. The condition for the resonance interaction (\ref{1}) can be satisfied for a sufficiently long time until the amplitude of the plasma  wave becomes higher than that of the laser incident wave and the reversed process of feeding back of the incident wave starts. If the laser is powerful enough than the instability will keep growing until a breakdown (overturn) of the plasma wave occurs, i.e. when the magnitude of the saturation plasma wave amplitude becomes equal to ${E_p}_{max}$ (see Eq. \ref{44}).  
 The acceleration constraint (\ref{44}) enables us to control the acceleration process. In order to determine the electron acceleration time we need to estimate the duration of acceleration, i.e. the time interval during which the phase velocities of a plasma wave and that of an electron remains approximately equal to each other.  \\
	\indent Taking into account this condition we can estimate the duration of acceleration of a   relativistic electron with an initial energy $\varepsilon=mc^2 (\gamma-1)$, $\gamma=1/\sqrt{1-{u_0}^2/c^2}>>1$, $u_0$ is the speed of an electron beam. The speed of such an electron is the following: 

\begin{equation}
\label{5}
\frac{u_0}{c}=1- \frac{1}{2\gamma^2}
\end{equation}

 Using the following estimation formula for acceleration time:

\begin{equation}
\label{6}
\vert V_p-u_0\vert\tau_f \approxeq c\pi\sqrt{\gamma_0}/\omega_p.
\end{equation}

 and Eqs. (\ref{3}) and (\ref{5}) we can determine the acceleration time:

\begin{equation}
\label{7}
\tau_f \approxeq \frac{2\pi{\gamma_0}^{3/2}\gamma^2\omega_0^2}{\omega_p\vert \omega_p^2\gamma^2-\omega_0^2\gamma_0\vert.}
\end{equation}
At $\gamma=1$ this equation transforms into equation for the acceleration of electron at a rest by  the plasma wave generated by the forward-scattering: $\tau_f=2\pi\sqrt{\gamma_0}/\omega_p$, which is much less than that for the relativistic electron at $\gamma>>\sqrt{\gamma_0}\omega_0/\omega_p$:  $\tau_f=2\pi {\omega_0}^2{\gamma_0}^{3/2}/{\omega_p}^3$ $\;\;$\cite{Dawson2}.\\
Taking into account Eq. (\ref{4}) and (\ref{7}) the following momentum and energy growth can be  obtained:
\begin{equation}\label{8}
\Delta P\approx e {E_p}_{max} \tau_f, \:\:\:\:\: \Delta\varepsilon \approx e {E_p}_{max} \tau_f c.
\end{equation}
\indent Let us make some estimations for a laser-driven plasma-based electron accelerator and compare with the results obtained at the 
\texttt{SPARC\char`_LAB} facility of INFN-LNF in Frascati, Italy \cite{Ross} where we presume that the stimulated forward-scattering scheme employing the relativistic electrons was realized, i.e. acceleration of injected relativistic electrons to the higher energies.

\begin{itemize}
\item {\textbf{Estimated parameters}}
\item[-]  $\omega_0=2.35\cdot 10^{15}$ s$^{-1}$ ($\lambda=800 $ nm), $\omega_p=1.8\cdot 10^{13}$ s$^{-1}$, $n_e=10^{17}$ cm$^{-3}$, $I= 10^{20}$ W/cm$^2$, $E_0\simeq 0.3$ TeV, $\gamma_0\simeq 7$.
\item[-] An electron with energy growth of $150$ MeV ($\gamma=297$) can gain the maximum energy $\Delta\varepsilon\simeq 1.6 $ TeV in the  breakdown field of ${E_p}_{max}\simeq 1 $ GV/cm during $\tau_f=50$ ns over the maximum acceleration length of $L \simeq 14$ m. Provided that the length of a capillary could be of the same size order: $L \simeq 14$ m, the corresponding energy growth for the \texttt{SPARC\char`_LAB} facility can be estimated to much higher energy growth of $\Delta\varepsilon\simeq 70 $ GeV compared to the obtained one inside the 8 cm-length capillary: $\sim 0.40$ GeV. Comparing our results with the results obtained by the \texttt{SPARC\char`_LAB}  for the same capillary length: $L \simeq 8$  cm, we will get the energy growth of $9$ GeV which is one order higher than the obtained one in the experiment. This is due to the missing data, i.e. we do not know the exact laser pulse intensity and the electron beam current used in the experiment.
\item[-] The acceleration time in a frame of  the stimulated backward-scattered model can be determined as $\tau_b= \pi/(2\omega_0)$, $\tau_b \simeq 10^{-15}$ s which is much less than that for the forward-scattered case: $\tau_f\simeq 50\cdot10^{-9}$ s by approx. $(\omega_0\sqrt{\gamma_0}/\omega_p)^3$ times (provided that  $\gamma>> \omega_0\sqrt{\gamma_0}/\omega_p$). The corresponding plasma wave lengths will be $\lambda_b= \pi c/\omega_0$, $\lambda_b\simeq 0.5 \mu  $m and $\lambda_f= 2\pi c\sqrt{\gamma_0}/\omega_p$, $\lambda_f= 280 \mu$m. We must note that account of the relativistic velocities of plasma electrons leads to the longer acceleration time by ${\gamma_0}^{3/2}$ times.
\end{itemize}
\indent In conclusion, we would like to note that an attempt to realize the acceleration scheme for  the stimulated forward-scattering model in a dense plasma ($n_e\sim 1.5\cdot 10^{19}$ cm$^{-3}$) was made in \cite{Modena}. However, we presume that there is no sufficient evidence for realization of such an experiment, since only the plasma electrons were accelerated with the low initial energy, whereas the prerequisite is the high electron relativistic injection energy in order to be in an  acceleration phase with the plasma wave of a relativistic phase velocity. Moreover, due to the high plasma density the plasma wave phase velocity was low what leads to breaking of the parametric resonance condition. This led to the low acceleration efficiency of $\sim 40$ MeV which is quite low compared to ours of $\sim TeV$. \\
\indent In the present work for the first time a new approach to solution of the analytical problem of relativistic interaction of a high-intensity laser pulse with  plasmas in a frame of a stimulated forward-scattering based electrons acceleration model has been proposed. We must note that the relativity of the beam electrons is crucial for our here proposed acceleration scheme.\\
 \indent The acceleration scheme employing the stimulated backward-scattered wave for particle acceleration in a wakefield is not suitable for particle acceleration  because high-intensity lasers  have a very short laser pulse length leading to a very short interaction time between the injected  electron beam and the plasma wave. Instead, our new approach employing the stimulated  forward-scattering and injection of relativistic electrons can provide more durable particle acceleration time inside the field of a plasma wave of a much longer length compared to the backward-scattered model where electrons just slip off the wave. For such a realization the injected electrons should be relativistic. The relativity of plasma electron velocities assists at the better convergence of the phase velocities of the plasma wave and of the laser with that of the  beam electrons, respectively with the speed of light, providing the longer acceleration phase. In Eq. for a breakdown electric field for illustration purpose we took a constant topf field because the exact electric field profile during the instability growth is not known. However, the time interval during which the breakdown field can be gained is much less than the acceleration time what justifies our choice of the field profile. \\
\indent To assess how a forward-scattering based scheme compares quantitatively with the laser-driven wakefield acceleration scheme employing a ponderomotive force we are planning to run  extensive simulations of the considered phenomena using available PIC codes. 
\\
\\
 S.P. Sadykova would like to express her gratitude to the Helmholtz Foundation for its financial support of the work.

\appendix

\nocite{*}

\end{document}